\documentclass[aps,prd,twocolumn,superscriptaddress,showpacs,amsmath,amssymb,amsthm,nofootinbib]{revtex4-2}
 \usepackage{comment}
\usepackage{amsmath,amssymb}
\usepackage{enumitem}
\usepackage{ulem}
\usepackage{mathrsfs}
\usepackage[usenames, dvipsnames]{color}
\usepackage{graphicx}
\usepackage{subcaption}
\usepackage[utf8]{inputenc}
\usepackage{url}
 \usepackage{xcolor}

\usepackage{hyperref}
\usepackage{float}
\usepackage{graphicx}
\usepackage{dcolumn}
\usepackage{bm}

\newcommand{\be}{\begin{equation}}             
\newcommand{\ee}{\end{equation}}               
\newcommand{\ba}{\begin{eqnarray}}
\newcommand{\ea}{\end{eqnarray}}

\begin{document}

\title{Mass inflation and thin shells in quasi-topological gravity
}

\author{Francesco Di Filippo}
\email{difilippo@itp.uni-frankfurt.de}
\affiliation{Institut f\"ur Theoretische Physik, Max-von-Laue-Str. 1, 60438 Frankfurt, Germany}
\author{David Kubiz\v n\'ak}
\email{david.kubiznak@matfyz.cuni.cz}
\affiliation{Institute of Theoretical Physics, Faculty of Mathematics and Physics, 
 Charles University, V Hole\v{s}ovi\v{c}k\'{a}ch 2, 180 00 Prague 8, Czech Republic}
\author{Aravindhan Srinivasan}
\email{srinivasan@math.cas.cz}

\affiliation{Institute of Theoretical Physics, Faculty of Mathematics and Physics, 
 Charles University, V Hole\v{s}ovi\v{c}k\'{a}ch 2, 180 00 Prague 8, Czech Republic}

\affiliation{ Institute of Mathematics, Czech Academy of Sciences, \v Zitn\'a 25, 115 67 Prague 1, Czech Republic}

\date{\today}

\begin{abstract}

We study the null junction conditions in  (re-summed) quasi-topological gravity theories, showing that no null thin shells exist within the realms of standard distributional theory for the 
pure gravity regular black hole solutions we have analyzed. This implies that the usual derivation of the mass inflation instability, which makes use of null thin shells, is not applicable in these theories. The problem of stability of 
inner horizons of 
regular black holes in quasi-topological gravity is hence still open and must be addressed with a more refined analysis, which does not rely on thin shells or the vacuum condition. 

\end{abstract}

\maketitle

\section{Introduction}
\label{section_intro}
{\it Regular black holes} are a {conservative} solution of the singularity problem in which a central singularity is replaced by a regular core shielded by an inner horizon. Such a geometry can mimic extremely well the properties of {\it General Relativity} (GR) black holes -- thus passing all possible observational tests. However, the simplicity of the solution is hindered by potential self-consistency issues. In fact, the presence of an inner horizon can lead to an instability due to the phenomenon of {\it mass inflation} which consists of an exponential blueshift amplification of small perturbation \cite{Poisson:1990eh,Ori:1991zz,Carballo-Rubio:2018pmi,Carballo-Rubio:2021bpr,Carballo-Rubio:2024dca}.

Remarkably, assuming the presence of null thin shells, this instability can be derived from a geometrical analysis supplemented by minimal dynamical assumptions 
\cite{Carballo-Rubio:2018pmi,DiFilippo:2022qkl}. 
The simplest perturbative setup exhibiting mass inflation consists of two null shells following geodesics of the background spacetime and intersecting on a co-dimension two hyper-surface located arbitrarily close to the inner horizon. In GR, when this configuration is considered for black holes possessing an inner horizon, the {Misner--Sharp mass} {\it grows  exponentially} as the intersection surface approaches the horizon \cite{Poisson:1990eh}.

For regular black holes, an analogous result can be obtained under the assumption that the discontinuity of the Misner--Sharp mass across a null shell is proportional to the shell’s energy~\cite{DiFilippo:2022qkl,Carballo-Rubio:2022kad}. This, together with the existence of thin shells, constitutes the only input from the {\it dynamical} equations of a given theory, while the rest of the analysis is purely geometrical and therefore applies to any geometrical theory of gravity. This suggests that mass inflation is a highly {\it universal phenomenon}, expected to arise in a broad class of theories in a manner largely independent of the specific underlying dynamics, provided the thin shell formalism is {\it well-defined} in the theory at hand.\footnote{Note that the (fine tuned) inner extremal black holes may be exempt from such an  instability, e.g. \cite{Colleaux:2019ckh, Carballo-Rubio:2022kad, DiFilippo:2024mwm}. }

Recently, regular black hole solutions have been obtained as vacuum solutions within a specific class of modified gravity theories known as (re-summed) {\it Quasi-Topological (QT)} gravity \cite{Colleaux:2019ckh, Bueno:2019ycr, Bueno:2024dgm,Bueno:2024eig, Bueno:2024zsx}. This development provides a natural framework in which to test the persistence of the mass inflation instability beyond GR. In fact, according to the recent study  
\cite{Frolov:2026rcm}, coming from purely kinematic considerations, the mass inflation was found to be `tamed' for the regular black hole spacetimes that are vacuum solutions in this class of theories.\footnote{See also \cite{Liu:2026xqf, Liu:2026ltw} for some recent studies on regular black holes, motivated from a loop quantum gravity (LQG) viewpoint. In particular, \cite{Liu:2026ltw} carries out a (kinematical) mass inflation analysis of LQG-inspired regular black holes, akin to \cite{Frolov:2026rcm}.}

In this paper, we aim to  
re-investigate this issue going beyond kinematical considerations by taking into account the dynamics of the QT theories of gravity.
To this end,
we attempt to derive the junction conditions for null shells in QT gravity. We employ several different prescriptions, obtaining, in particular,   
a generic constraint that must be satisfied for the equations of motion to be defined in the distributional sense. We study this constraint for several examples of QT theories and their corresponding vacuum solutions. Interestingly, we find that for the Gauss--Bonnet and pure Lovelock theories \cite{Lovelock:1971yv,Dadhich:2012cv, Gannouji:2019gnb} the junction conditions allow for null thin shells. However, for the examples of regular black holes in QT theories, the obtained junction conditions forbid null thin shells, hinting on the fact that the thin shell formalism may not be well defined for such theories. 

Our paper is organized as follows. In Sec.~\ref{section_Double_null}, we generalize the double null shell formalism \cite{Barrabes1990, Barrabes:1991ng, Carballo-Rubio:2018pmi,DiFilippo:2022qkl} to an arbitrary number of dimensions and introduce theories of QT gravity in Sec.~\ref{sec:QTtheories}. In Sec.~\ref{Sec:3}, the junction conditions for null shells in QT gravity are derived. 
These are then applied to various static solutions in (special) Lovelock gravities in Sec.~\ref{Love_Lock}, and to various (vacuum) regular black holes in QT gravity in Sec.~\ref{Sec:4}, showing that null thin shells lead to inconsistencies and cannot be present in those regular black hole spacetimes.
We conclude in Sec.~\ref{Sec:5}.
In Appendix~\ref{Append_1}, we provide the technical details of a result discussed in Sec.~\ref{subsection_second_jn}, and in Appendix~\ref{mass_inflation_RN_app}, we review mass inflation in the $D$-dimensional charged Schwarzschild--Tangherlini black hole \cite{Tangherlini:1963bw}, specifically using the junction conditions derived in the main text.


\section{Double null shell formalism in $D$ dimensions} \label{section_Double_null}
Consider a static spherically symmetric spacetime in $D \geq 4$ dimensions. In an Eddington--Finkelstein-like coordinate system, its metric has the following general form: 
\ba 
ds^2&=&g_{ab}dx^a dx^b\nonumber\\
&=&- e^{-2\chi (r)} F(r) dv^2 + 2 e^{-\chi(r)} dv dr + r^2 d\Omega^2_{D-2}\,,\quad \label{null_coord_metric}
\ea
where $a,b, \dots$ denote the indices on the $D$-dimensional spacetime, $d\Omega_{D-2}^2$ is the line element on the $(D-2)$-dimensional unit sphere, and $F$ and $\chi$ are metric functions determined by dynamics. Let us also define the indices $i,j,\dots$ as those on the $(D-2)$-sphere, so that they take values in $2, \dots, (D-1)$. We are interested in regular black holes. Therefore, $F$ must have an even number of zeros, which we can assume to be two for simplicity~\cite{Hayward:2005gi,Carballo-Rubio:2018pmi}; in the presence of extra horizons the analysis can be repeated focusing only on the two outermost ones.

\begin{figure}
    \centering
\includegraphics[width=0.5\linewidth]{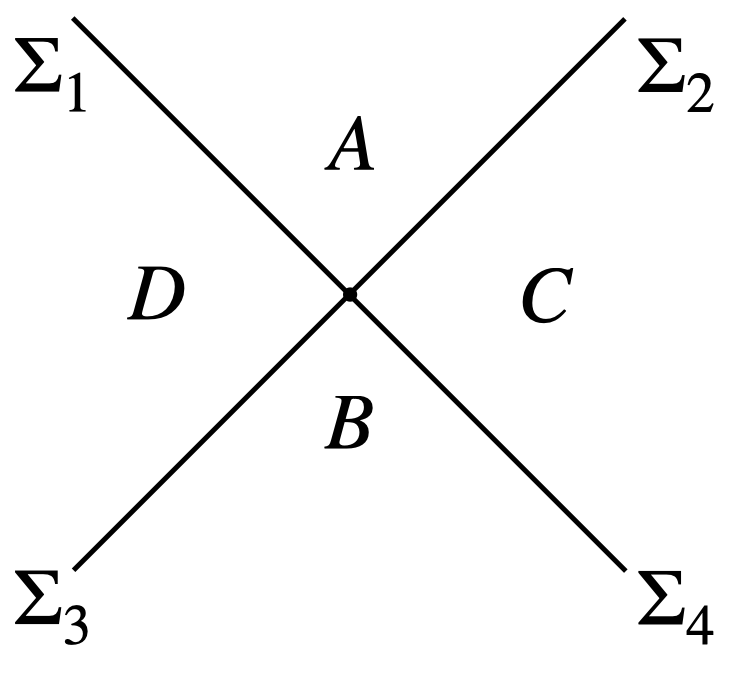}
    \caption{A pair of spherical null shells crossing at $r_0$ and dividing the spacetime into four regions.}
    \label{fig:Double_null_shells}
\end{figure}

We perturb this background by a pair of spherically symmetric null thin shells \cite{Barrabes1990,Barrabes:1991ng,Carballo-Rubio:2018pmi,DiFilippo:2022qkl}. As depicted in Fig.~\ref{fig:Double_null_shells}, the shell surfaces $\Sigma_3$ and $\Sigma_4$ intersect on the $(D-2)$-dimensional hyper-sphere $S$ and re-emerge as the surfaces $\Sigma_1$ and $\Sigma_2$, dividing the spacetime into four regions: $A$, $B$, $C$ and $D$. Let us assume the existence of a coordinate system that continuously extends across the four regions in a neighborhood of $S$ (and is piecewise differentiable \cite{DiFilippo:2022qkl}), and that in these coordinates the radius of $S$ is $r_0$.

A $(D-2)$-dimensional spatial surface in $D$ dimensions has exactly two null directions orthogonal to it. Therefore, on the hypersphere $S$, the null normals of $\Sigma_1$ and $\Sigma_4$ must be aligned, and likewise those of $\Sigma_2$ and $\Sigma_3$, i.e.,
\begin{align}
l_1^{a}=\alpha l_4^{a}\,, \quad l_2^a=\beta l_3^a\,, \quad (\text{evaluated on } S)\ \label{null_normals_constraint}
\end{align}
where $l_1, \dots, l_4$ respectively denote the null normals of the surfaces $\Sigma_1,\dots , \Sigma_4$. Note that the relation \eqref{null_normals_constraint} holds only in a continuously extending coordinate system and can be appropriately contracted to obtain the following coordinate-independent relation:
\begin{align}
(l_{(1)}.l_{(2)})_A (l_{(3)}.l_{(4)})_B= (l_{(1)}.l_{(3)})_D (l_{(2)}.l_{(4)})_C\,,\label{contracted_null_rels}
\end{align}
where the subscripts $A, \dots, D$ denote the region where each contraction is evaluated. 

Let us denote the extrinsic curvatures of the surfaces $\Sigma_1, \dots, \Sigma_4$, respectively, by $K^{(1)}_{ij}, \dots, K^{(4)}_{ij}$. For each of these, one can define the scalar $K=\sigma^{ij}K_{ij}$, where $\sigma_{ij}$ is the induced metric on the corresponding surface. Assuming that the induced metrics extend continuously across the two sides of each surface, one finds that the extrinsic curvatures and hence the associated scalars are also continuous \cite{poisson2004relativist}. Therefore, equation \eqref{contracted_null_rels} leads to \cite{DiFilippo:2022qkl}
\begin{align}
   & \left(\frac{K^{\mbox{\tiny (1)}}K^{\mbox{\tiny(2)}}}{ l_{\mbox{\tiny(1)}}.l_{\mbox{\tiny {(2)}}}}\right)_{\mbox{\tiny A} }\left(\frac{K^{\mbox{\tiny(3)}}K^{\mbox{\tiny (4)}}}{ l_{\mbox{\tiny (3)}}.l_{\mbox{\tiny (4)}}}\right)_{\mbox{\tiny B}}=  \left(\frac{K^{\mbox{\tiny (1)}}K^{\mbox{\tiny(3)}}}{ l_{\mbox{\tiny(1)}}.l_{\mbox{\tiny {(3)}}}}\right)_{\mbox{\tiny D} }\left(\frac{K^{\mbox{\tiny(2)}}K^{\mbox{\tiny (4)}}}{ l_{\mbox{\tiny (2)}}.l_{\mbox{\tiny (4)}}}\right)_{\mbox{\tiny C}}\!\!. \label{extr_prod}
\end{align}

The explicit expression for $K$ of a surface can be evaluated using its null normal $l$ as \cite{poisson2004relativist,DiFilippo:2022qkl}
\ba
K&=& \sigma^{ij}K_{ij}=  \sigma^{ij}\mathcal{L}_{l} \sigma_{ij}\nonumber\\
&=&\sigma^{ij}l^{c}\partial_{c}\sigma_{ij}= \frac{2(D-2)}{r}l^{a}\partial_{a}r\,, \label{extr_scalar}
\ea
where $\mathcal{L}$ is the Lie derivative, and we have used the canonical form of $\sigma_{ij}$ (read off from the ansatz \eqref{null_coord_metric}).

Assuming that a Birkhoff theorem holds in the theories of interest and that  $r$ and the angular coordinates extend continuously, implies that the metrics in the four regions are static and take exactly the form given by \eqref{null_coord_metric}, possibly with different functions $F$ and $\chi$, and different $v$ coordinates. Therefore, proceeding analogously to \cite{DiFilippo:2022qkl}, i.e., using equation \eqref{extr_scalar} and the completeness relations satisfied by the metrics in each region (see relations (10) of \cite{DiFilippo:2022qkl}), equation \eqref{extr_prod} leads to the well-known {\it Dray--t'Hooft--Redmount (DTR)} relation \cite{Dray:1985yt,1985PThPh..73.1401R}, now explicitly derived in $D$ dimensions:\footnote{The generalization of the DTR relation to $D$ dimensions was noted in \cite{Frolov:2026rcm} without explicit derivation.}
\begin{align}
F_A(r_0)F_B(r_0)=F_C(r_0)F_D(r_0) \quad (\text{evaluated on } S)\label{DTR}\,.
\end{align}
In terms of the Misner--Sharp quasi-local mass \cite{Misner:1964je, Hernandez:1966zia}\footnote{Although not relevant for our purpose, let us point out that there exist generalizations of the Misner–Sharp mass to modified theories of gravity, e.g. \cite{Maeda:2006pm,Maeda:2007uu,Cai:2009qf}.}
\begin{align}
    M(r)\equiv\frac{1}{2}r^{D-3}\Bigl(1-F(r)\Bigr)\,, \label{MS_mass}
\end{align}
the DTR relation \eqref{DTR} can be rewritten as
\begin{align}
     M_A(r_0)&= M_B(r_0)+ M_{in}(r_0) + M_{out}(r_0) \nonumber\\
     &-\frac{2 M_{in}(r_0) M_{out}(r_0)}{r^{D-3}_0F_B(r_0)}\,, \label{Mass_rels}
\end{align}
where the quantities $M_{in}$ and $M_{out}$, defined below, are interpreted as the Misner--Sharp masses associated with the ingoing and outgoing shells \cite{Carballo-Rubio:2018pmi,DiFilippo:2022qkl}:
\ba
   M_{in}(r_0)&\equiv& M_{C}(r_0)-M_B(r_0)\,,\label{Min} \nonumber\\
   M_{out}(r_0)&\equiv& M_{D}(r_0)-M_B(r_0)\,. \label{M_out}
\ea

The relation \eqref{Mass_rels} is the key ingredient in the analysis of mass inflation \cite{Barrabes1990}, where one determines how the two perturbing shells affect the region $A$, when the ingoing null shell $\Sigma_4$ is arbitrary close to the inner horizon, so that \mbox{$r_0\rightarrow r_-$}. To do this, we first examine the motion of the outgoing shell $\Sigma_3$ in region $B$, given by
\begin{align}
\frac{dr}{dv_{B}}= \frac{e^{-\chi_B}}{2}F_B(r)\,.\label{sigma_3_motion}
\end{align}
Evaluating \eqref{sigma_3_motion} on $S$ and expanding $r_0$ near $r_-$, we get
\begin{align}
 \delta r \sim e^{-|\kappa_{-}|v_B}\,, \label{delta r_eqn}
\end{align}
where $\delta r = r_0-r_-$, and we have used the definition of surface gravity on the inner horizon
\begin{align}
 \kappa_{-} = -|\kappa_{-}|=\Big(\frac{e^{-\chi}}{2}F'(r)\Big)\Big|_{r=r{-}}\,,   
\end{align} 
along with the expansion of $F_B(r_0)$ near $r_-$
\begin{align}
    F_B(r_0)= F_B'(r_-) \delta r +  O(\delta r^2)\,. \label{F_expansion}
\end{align}
If we plug  \eqref{delta r_eqn} back into the expression for $F_B(r_0)$, we obtain
\begin{align}
    F_B(r_0) \sim F'_B(r_-)e^{-|\kappa_{-}|v_B}\,.\label{F_Beqn}
\end{align}
Therefore, substituting the $v$-dependence of $F_B$ into \eqref{Mass_rels} leads to
\begin{align}
     M_A(r_0)=& M_B(r_0)+ M_{in}(r_0) + M_{out}(r_0) \nonumber\\
     &-\frac{2e^{|\kappa_{-}|v_B} M_{in}(r_0) M_{out}(r_0)}{r^{D-3}_0F'_B(r_-)}\,.\label{Mass_rels_2}
\end{align}
Note that equation \eqref{Mass_rels}, or equivalently \eqref{Mass_rels_2}, is purely kinematical. Furthermore, we see from equation \eqref{delta r_eqn} that \mbox{$r_0\rightarrow r_-$} implies \mbox{$v_B\rightarrow \infty$}. Therefore, when the trajectory of the ingoing shell is taken close to that of the inner horizon, the Misner--Sharp mass $M_A$ contains an exponentially diverging factor. Nevertheless, to determine whether $M_A$ indeed diverges, one also has to determine the $v$-dependence of all the other terms on the right hand side of \eqref{Mass_rels_2}, which requires dynamics-dependent junction conditions for null shells. Moreover, although purely geometric, the validity of the double null shell formalism requires the existence of null thin shells in the first place, which in turn also depends on the junction conditions.\footnote{Previous works on the mass inflation of regular black holes (see e.g. \cite{Carballo-Rubio:2018pmi}) assume the existence of null thin shells, and hence the validity of \eqref{Mass_rels_2}. Furthermore, they assume that $M_{in}$ and $M_{out}$ are proportional to the corresponding shell mass, leading to an exponential mass inflation instability, as in GR (see Appendix \ref{mass_inflation_RN_app}).} 
Therefore, to analyze the mass inflation in regular black hole spacetimes, it becomes necessary to derive the junction conditions in the corresponding theories -- the QT gravity in our case.

\section{Theories of QT gravity}\label{sec:QTtheories}

QT gravity is a class of theories in $D\geq 5$, forming a generalization of Lovelock theories \cite{Lovelock:1971yv}, defined as those with second order equations of motion when restricted to static spherically symmetric metrics \cite{Bueno:2019ycr, Bueno:2022res}. As these theories have degeneracies at the level of the action \cite{Moreno:2023rfl}, i.e., there are different QT Lagrangians that yield the same equations of motion on static spherically symmetric metrics, the papers \cite{Bueno:2024zsx, Bueno:2024eig} study a further restriction of these theories. The subclass is defined by relaxing the staticity condition and demanding that the equations of motion of the theories are at most second order on general spherically symmetric metrics.

The Lagrangians of these theories are identified by working with the following ansatz \cite{Bueno:2024zsx, Bueno:2024eig}:\footnote{The Lagrangians identified in \cite{Bueno:2024zsx, Bueno:2024eig} are symmetry-reduced Lagrangians, meaning that the spherical symmetry is already imposed at the level of the action rather than on the equations of motion. Although this is a well-known trick to obtain simplified equations of motion \cite{Weyl:1917gp, Lovelock1973SphericallySM}, its justification requires certain conditions to be satisfied so that the variation of the action commutes with the symmetry reduction \cite{Frausto:2024egp}.}
\begin{align}
   ds^2=g_{ab}dx^a dx^b  = h_{\mu \nu }dx^\mu dx^\nu  + \varphi(x)^2 d\Omega^2_{D-2}\,, \label{dil_grav_ansatz}
\end{align}
 which uses spherical symmetry to split the full metric $g_{ab}$ into a two-dimensional metric $h_{\mu \nu}$ ($\mu, \nu, \dots=0,1$) and a sphere whose radius is governed by the dilaton field $\varphi$, both of which depend only on the coordinates $x^\mu$ defined on the two-dimensional part. 
The resultant equations of motion are given by \cite{Bueno:2024zsx}
\ba
  {8\pi G_{\!N} T_{\mu\nu}=\mathcal{E}_{\mu \nu }}\!\!&=&\!\!\frac{D\!-\!2}{\varphi^{D\!-\!2}}\Big(G_3 h_{\mu [\nu} \nabla_{\beta]}\partial^\beta \varphi -\frac{G_2}{2}h_{\mu \nu} \Big),\quad \ \label{E_mu_nu}\\
   {8\pi G_{\!N}T_{ij}}=\mathcal{E}_{ij}\!\!&=&\!\!\frac{1}{2\varphi^{D-3}}\Big( \Box\varphi G_{3,\varphi}- G_{2,\varphi}-\frac{G_3}{2}R\qquad\nonumber\\
   &&-2 \nabla_{[\rho}\partial^\sigma\varphi \nabla_{\sigma]}\partial^\rho \varphi G_{3,X}\Big)g_{ij}\,,\label{E_ij}
\ea
where $\nabla_\mu$ is the covariant derivative associated with $h_{\mu \nu}$, and $R$ is the corresponding two-dimensional Ricci scalar, ${\cal E}_{ab}$ is the generalized Einstein tensor associated with the full geometry, {and $T_{ab}$ is the matter energy-momentum tensor.} The angular components of the metric are defined by \mbox{$g_{ij}dx^i dx^j = \varphi^2 d\Omega^2_{D-2}$}, and the definitions of the functions in \eqref{E_mu_nu} and \eqref{E_ij} read
\begin{align}
   & X= \nabla_\mu\varphi \nabla^\mu \varphi\,, \quad \psi=\frac{1-X}{\varphi^2}\,,  \label{X_defn}\\
   & h(\psi)=\psi + \sum_{n=2}^{\infty} \alpha_n \frac{D-2n}{D-2}\psi^n\,, \label{hpsi}\\
    & G_{2}(\varphi, X)= \varphi^{D-2}[(D-1)h(\psi)-2\psi h'(\psi)]\,,\label{G_2_defn}\\
    & G_3(\varphi, X)=2 \varphi^{D-3}h'(\psi)\,,\label{G_3_defn}
\end{align}
where $\alpha_n$ are the coupling constants of the higher-curvature terms, the prime denotes a derivative with respect to $\psi$, and $G_{,X}$ and $G_{,\varphi}$ denote the partial derivatives of a function $G$ with respect to $X$ and $\varphi$, respectively.

Upon gauge fixing $\varphi=r$ and choosing the two-dimensional metric as
\ba
    d\gamma^2&=&h_{\mu \nu}dx^\mu dx^\nu
    \nonumber\\
    &=&- e^{-2\chi (v,r)} F(v,r) dv^2 + 2 e^{-\chi(v,r)} dv dr\,,
\ea
it is shown that these theories satisfy a Birkhoff theorem, i.e., vacuum solutions of equations \eqref{E_mu_nu}, \eqref{E_ij} are automatically static and hence reduce to the form \eqref{null_coord_metric}. Therefore, these are referred to as Birkhoff-QT theories \cite{Bueno:2024zsx, Bueno:2024eig} (see also \cite{Bueno:2025qjk} for a more recent classification). 

In this gauge, the expression for $\psi$ becomes
\begin{align}
     \psi=\frac{1-F(r)}{r^2}= \frac{2M(r)}{r^{D-1}}\,,\label{psi_F_rel}
\end{align}
where we have used the definition of the Misner--Sharp mass \eqref{MS_mass}. Furthermore, the vacuum equations fix $\chi=0$ and also determine $h(\psi)$ as \cite{Bueno:2024zsx, Bueno:2024eig}
\begin{align}
h(\psi)=\frac{2m}{r^{D-1}},
\label{h_psi_soln}
\end{align}
where $m$ is an integration constant.

Our study in this paper will focus on the Birkhoff-QT class. In particular, we will derive junction conditions for spherically symmetric null shells in these theories using equations of motion \eqref{E_mu_nu}, \eqref{E_ij} by working with the general spherically symmetric ansatz \eqref{dil_grav_ansatz}, without demanding staticity or the vacuum condition. Once we derive the junction conditions applicable to the more general case, we will then specialize to static spherically symmetric spacetimes, which includes the charged Schwarzschild--Tangherlini and vacuum regular black holes.

\section{Junction conditions for a null hypersurface in spherically symmetric solutions of QT gravity}\label{Sec:3}

\subsection{Geometric setup}

Following the standard approach \cite{poisson2004relativist}, we have the null hypersurface, denoted by $\Sigma$, splitting the regions of spacetime into $\mathcal{M}_-$ and $\mathcal{M}_+$, with $+$ denoting the future region and $-$ the past. We assume that $\Sigma$ preserves spherical symmetry and thus define the coordinates $x^a_{\pm}= (x^\mu_{\pm}, \theta^i)$ in the respective regions such that the angular coordinates $\theta^i$ are the same on both sides of the surface. In these coordinates of the respective regions, the $2D$ metric and the dilaton are $h^{\pm}_{\mu \nu }$ and $\varphi^{\pm}$. Let the parameter $\lambda$ parameterize the normal null curves to $\Sigma$ on both sides of the surface, so that we have the coordinates $(\lambda, \theta^i)$ on $\Sigma$. We can thus define the null vector $k^a_{\pm}$ and the spatial tangent vectors $e^{a}_{\pm i}$ as
\begin{align}
    k_{\pm}^a= \Big(\frac{\partial x_{\pm}^a}{\partial \lambda}\Big )_{\theta^i}\,, \quad e^{a}_{ \pm i}=\Big(\frac{\partial x^a_{\pm}}{\partial \theta^i}\Big )_{\lambda}\,.
\end{align}
By definition, we have the following relations satisfied by the vectors on their corresponding sides
\begin{align}
    k^a k_a=0=k_a e^{a}_{i}\,.
\end{align}
Considering the splitting of $x_{\pm}^a$ into $2D$ and the spherical parts, we have the following: 
\begin{align}
   &k_{\pm}^i=0\,, \quad e^{j}_{ \pm i}=\delta_{ij}\,, \quad e^{\mu}_{ \pm i}=0\,. 
\end{align}
Therefore, the induced metric $\sigma_{ij}$ on $\Sigma$ simply becomes
\begin{align}
    \sigma^{\pm}_{ij}=g^{\pm}_{ab}e^{a}_{i}e^{b}_{j}= g_{ij}^{\pm}\,,
\end{align}
and the corresponding line element reads
\begin{align}
    ds^2_{\pm\Sigma}= \sigma^{\pm}_{ij}d\theta^i d\theta^j= g^{\pm}_{ij}d\theta^i d\theta^j= \varphi^{2}_{\pm } d\Omega^2_{D-2}\,.\label{line_elt_Sigma}
\end{align}
It is then conventional to demand that the line elements (or equivalently the induced metrics) on the two sides be the same. Hence, from equation \eqref{line_elt_Sigma}, we have the following:
\begin{align}
[\varphi]_{\pm}=0\,,\label{phi_cont}
\end{align}
where $[Y]_{\pm}\equiv Y^+- Y^-$ is evaluated on $\Sigma$. 

On each side, let us also introduce the standard auxiliary null vectors $N^{a}_{\pm}$ adapted to the spherical coordinates, i.e., $N^{a}_{\pm}=(N^\mu_{\pm}, 0)$, which are obviously orthogonal to $e_{i}^{a}$ and satisfy the following on their respective sides \cite{poisson2004relativist}:
\begin{align}
N_{\mu} N^{\mu}=0\,, \quad N^\mu k_{\mu}=-1\,.\label{null_aux_defn}
\end{align}
Thus, on each side, we have the following completeness relation:
\begin{align}
h_{\mu \nu}=-k_\mu N_\nu-N_\mu k_\nu\, .\label{completeness_2D}
\end{align}

Finally, let us assume the existence of continuous coordinates $x^{\mu}$ (not necessarily the same as $x_{\pm}^{\mu}$) in a neighborhood of $\Sigma$. Furthermore, let us introduce a timelike geodesic congruence that intersects the null hypersurface. Let us choose a continuous parameter $\tau(x^{\mu})$ such that $\tau<0$, $\tau>0$ respectively correspond to the regions $\mathcal{M}_-$ and $\mathcal{M}_+$, with the intersection occurring at $\tau=0$. The tangent to these timelike curves, denoted by $u^{a}$, is given by \cite{poisson2004relativist}
\begin{align}
   u^\mu= \frac{dx^\mu}{d\tau}\,, \quad u^i=0\,.
\end{align}
Moreover, we can write $k_\mu$ as
\begin{align}
k_\mu= -(-k_\nu u^\nu) \frac{\partial \tau }{\partial x^\mu}\,,\label{k_rel_cont_coord}
\end{align}
As a consequence of the continuity of $x^\mu$ and $\tau$, we have 
\begin{align}
[u^\mu]_{\pm}=0\,, \quad [k_\mu]_{\pm}=0\,. \label{cont_k}
\end{align}

\subsection{First junction conditions}
Since the equations of motion \eqref{E_mu_nu}, \eqref{E_ij} of the Birkhoff-QT theories involve the fields $h_{\mu \nu}$ and $\varphi$, we can employ the methods of \cite{Aviles:2019xae} used in studying the junction conditions of scalar-tensor theories (see also \cite{Padilla:2012ze} for non-null boundaries in scalar-tensor theories; \cite{Senovilla:2013vra, Reina:2015gxa} on modified gravity theories; and \cite{Mars:1993mj,Senovilla:2018hrw} which obtain geometrical junction conditions for character changing hypersurfaces). As in \cite{Aviles:2019xae}, we define the following distributional notions for the $2D$ metric and the dilaton
\begin{align}
    & h_{\mu \nu }= h^+_{\mu \nu } \Theta(\tau)+h^-_{\mu \nu } \Theta(-\tau)\,, \label{h_mu_nu_distrb}\\
    & \varphi = \varphi^+ \Theta(\tau)+\varphi^- \Theta(-\tau)\,.\label{phi_dstrib}
\end{align}
Note that these are independent of the angular coordinates. Hence, we need to worry only about the derivatives with respect to $x^\mu$.

Taking the derivative of $h_{\mu \nu}$ and using equation \eqref{k_rel_cont_coord}, we obtain 
\begin{align}
   \partial_\rho h_{\mu \nu }=&  \partial_\rho h_{\mu \nu }^+ \Theta(\tau) + \partial_\rho h_{\mu \nu }^- \Theta(-\tau)\nonumber\\
  & -(-k_\beta u^\beta)^{-1}[h_{\mu \nu}]_{\pm}k_\rho\delta(\tau)\,.\label{dh}
\end{align}
As the equations of motion involve the Ricci scalar corresponding to $h_{\mu \nu}$, the Dirac delta term in \eqref{dh} would lead to products of deltas. Therefore, as is standard \cite{poisson2004relativist, Aviles:2019xae}, we remove the delta term by demanding
\begin{align}
    [h_{\mu \nu}]_{\pm}=0\,.\label{h_metric_continuity}
\end{align}
One obtains a similar expression for the dilaton field, where eventually the delta term can be eliminated by using the continuity \eqref{phi_cont}. Therefore, the derivatives of the $2D$ metric and the dilaton reduce to
\begin{align}
    &\partial_\mu \varphi=  \partial_\mu \varphi^+ \Theta(\tau) +  \partial_\mu \varphi^- \Theta(-\tau) \,,\label{dphi_2}\\
     &\partial_\rho h_{\mu \nu }=  \partial_\rho h_{\mu \nu }^+ \Theta(\tau) + \partial_\rho h_{\mu \nu }^- \Theta(-\tau)\,.\label{dh_2}
\end{align}

From the completeness relation \eqref{completeness_2D} and the continuity conditions \eqref{cont_k}, \eqref{h_metric_continuity}, one can also deduce that
\begin{align}
    [N^\mu]_{\pm}=0\,. \label{N_cont}
\end{align}
The continuity of $h_{\mu \nu}$ and $\varphi$ imply that their tangential derivatives are also continuous
\begin{align}
    &[\partial_\rho h_{\mu \nu}]_{\pm}k^\rho=0 =[\partial_\rho \varphi]_{\pm}k^\rho\,.
\end{align}
The non-trivial contributions to the jumps of the derivatives, which will be important for the second junction conditions, come from the projection along $N^\mu$
\begin{align}
   \gamma_{\mu \nu }\equiv [\partial_\rho h_{\mu \nu}]_{\pm}N^\rho &\quad \Leftrightarrow\quad   [\partial_\rho h_{\mu \nu}]_{\pm}=-k_\rho \gamma_{\mu \nu }\,,\label{dh_jump}\\
     W \equiv - [\partial_\rho \varphi]_{\pm}N^\rho &\quad \Leftrightarrow\quad [\partial_\rho \varphi]_{\pm}=Wk_\rho\, . \label{d phi_jump}
\end{align}

\subsection{Second junction conditions}
\label{subsection_second_jn}
For the second junction conditions, we need to evaluate the second derivatives of the $2D$ metric and the dilaton, for which we closely follow the steps in \cite{poisson2004relativist, Aviles:2019xae}, assuming the following standard relations:
\begin{align}
& \Theta(\tau)^2= \Theta(\tau)\,, \quad \Theta(\tau)\Theta(-\tau)=0\,,\label{Theta_prod}\\
   &\Theta(\tau) + \Theta(-\tau)=1\,. \label{Theta_identity}
\end{align}
For $\varphi$, we find
\begin{align}
    \nabla _\mu \nabla_\nu \varphi =&  \nabla _\mu \nabla_\nu \varphi^+ \Theta(\tau) + \nabla _\mu \nabla_\nu \varphi^- \Theta(-\tau)\nonumber\\
    &- (-k_\alpha u^\alpha)^{-1} k_\mu k_\nu W \delta(\tau)\,.\label{phi_2_der_distr}
\end{align}
For $h_{\mu \nu}$, from \eqref{dh_2} we have the following relation for the two-dimensional Christoffel symbols: 
\begin{align}
    \Gamma^\alpha_{\hspace{1mm}\beta \rho}= \Gamma^{+\alpha }_{\hspace{3mm}\beta \rho}\Theta(\tau)+  \Gamma^{-\alpha }_{\hspace{3mm}\beta \rho}\Theta(-\tau)\,.
\end{align}
It can be shown that the two-dimensional Riemann tensor then takes the following distributional form:
\begin{align}
    R^\alpha_{\hspace{1mm}\beta \gamma \delta}=&R^{+\alpha}_{\hspace{3mm}\beta \gamma \delta}\Theta(\tau)+R^{-\alpha}_{\hspace{3mm}\beta \gamma \delta}\Theta(-\tau)\nonumber\\
    &-(-k_\rho u^\rho)^{-1} \Bigl([\Gamma^\alpha_{\hspace{1mm}\beta \delta}]_{\pm}k_\gamma-[\Gamma^\alpha_{\hspace{1mm}\beta \gamma}]_{\pm}k_\delta\Bigr)\delta(\tau)\,,
\end{align}
where, using \eqref{dh_jump}, one can show that the discontinuity of the Christoffel symbols is 
\begin{align}
    [\Gamma^\alpha_{\hspace{1mm}\beta \gamma}]_{\pm}=-\frac{1}{2}\bigl( \gamma^{\alpha}_{\hspace{2mm}\beta}k_{\gamma}+\gamma^{\alpha}_{\hspace{2mm}\gamma}k_{\beta}-\gamma_{\beta \gamma}k^{\alpha}\bigr)\,. \label{christo_discont}
\end{align}
Hence, for the Ricci scalar, we obtain 
\begin{align}
    R&=  R^\alpha_{\hspace{1mm}\beta \alpha \delta}h^{\beta \delta}= R^\alpha_{\hspace{1mm}\beta \alpha \delta}h_+^{\beta \delta}\nonumber\\
    &=R^+\Theta(\tau)+ R^-\Theta(-\tau) +  (-k_\rho u^\rho)^{-1}\gamma_{\mu \nu}k^\mu k^\nu \delta(\tau)\,, \label{Ricci_scalar_distr}
\end{align}
where we have used the relations \eqref{Theta_identity} and \eqref{h_metric_continuity} to reduce $h^{\alpha \beta}=h_+^{\alpha \beta}\bigl(\Theta(\tau) +\Theta(-\tau)\bigr)= h^{\alpha \beta}_+$.
 Note that $\gamma_{\mu \nu}k^\mu k^\nu$ can be equivalently expressed as 
\begin{align}
\gamma_{\mu \nu}k^\mu k^\nu= 2[k^\mu k^\nu\nabla_{\mu }N_{\nu} ]_{\pm}\equiv 2[C_{\lambda\lambda}]_{\pm}\,.\label{C_lambda}
\end{align}

To obtain the second junction conditions, we need to assemble a few more ingredients, namely the distributional forms of the functions $G_2$ and $G_3$. From \eqref{dphi_2}, we have
\begin{align}
   X = X^+ \Theta (\tau)+X^- \Theta (-\tau)\,.\label{X_distrib}
\end{align}
Therefore, using equations \eqref{phi_cont} and \eqref{Theta_identity}, we obtain
\begin{align}
   & \psi= \frac{1-X}{\varphi^2}
= \psi^+ \Theta(\tau) + \psi^- \Theta(-\tau)\,.
\end{align}
Hence, we find
\ba
    h(\psi) &= &h (\psi^+) \Theta(\tau) +h (\psi^-) \Theta(-\tau)\,, \label{h_distr}\nonumber\\
        G_2(\varphi, X) &=& G_2(\varphi, X^+) \Theta(\tau)+G_2(\varphi, X^-) \Theta(-\tau)\,,\quad\label{G_dsitr}\\
    G_3(\varphi, X) &=& G_3(\varphi, X^+) \Theta(\tau)+G_3(\varphi, X^-) \Theta(-\tau)\,,\nonumber
\ea
and similar expressions for the derivatives of $G_2$, $G_3$ with respect to $\varphi$ and $X$.

Now we can try to extract the second junction conditions, i.e., the singular part of the energy-momentum tensor, by plugging in the distributional forms of each of the quantities in the equations of motion \eqref{E_mu_nu}, \eqref{E_ij}. However, in doing so, one encounters the $\Theta \delta$ terms. For example, consider the $\mathcal{E}_{\mu \nu}$ equation of motion \eqref{E_mu_nu}. The only term in these equations that contains a delta factor is $-\frac{D-2}{2\varphi^{D-2}}G_3 \nabla_{\mu}\partial_\nu \varphi$. Explicitly expanding each factor in terms of its distributional form, we find that the singular part of the $\mathcal{E}_{\mu \nu}$ equation reads
\begin{align}
 \frac{D-2}{2\varphi^{D-2}}[G_3^+ \Theta(\tau)+ G_3^-\Theta(-\tau)]   (-k_\alpha u^\alpha)^{-1}  k_\mu k_\nu W \delta(\tau)\, ,\label{no_shell_illus_intermediate}
\end{align}
which contains $\Theta \delta$ terms.

Since $\Theta \delta$ is not defined as a distribution, the standard GR derivations try to eliminate it \cite{poisson2004relativist}. Applying the same strategy in our case, we could demand 
\begin{align}
   G_3^+ = G_3^-, \label{G_3_cont}
\end{align}
so that, using \eqref{Theta_identity} we can write
\begin{align}
   \Big [ G_3^+ \Theta(\tau)+ G_3^-\Theta(-\tau)\Big ]=G_3^+\Big [  \Theta(\tau)+ \Theta(-\tau)\Big ]=G_3^+,
\end{align}
thus effectively avoiding the $\Theta \delta$ product, as in the evaluation of $R$ in \eqref{Ricci_scalar_distr}. From equation \eqref{G_3_defn}, we see that the condition $G_3^+ = G_3^-$ implies 
\be 
h'(\psi^+) = h'(\psi^-)\,.\label{cont_prescrpt}
\ee
This condition is consistent for GR, as GR is defined by \cite{Bueno:2024dgm, Bueno:2024zsx, Bueno:2024eig}
\begin{align}
h(\psi)=\psi , \label{h_GR}
\end{align}
hence $h'(\psi)=1$, automatically matching on both sides without additional constraints. However, for other theories in the Birkhoff-QT class, this can lead to non-trivial constraints on the shell. In particular, for the beyond-GR black holes discussed in this paper, we show in Appendix \ref{Append_1} that the condition \eqref{G_3_cont} forbids the existence of spherically symmetric thin shells, including non-null ones (see \cite{Senovilla:2026fby} for a discussion on the existence of thin shells in generic theories of gravity).

{In an attempt to allow for the existence of thin null shells,} 
we relax the continuity of $G_3$ and adopt one of the following two prescriptions to handle $\Theta \delta$ terms:\footnote{A rigorous treatment of such products requires the use of generalized (Colombeau) distributions \cite{Colombeau:1985,Colombeau:1992,gsponer2009concise,Silva:2026jiv}, which is beyond the scope of this paper.}
\begin{enumerate}
    \item \label{prescrip_1} In the equations of motion, we first apply the rules \eqref{Theta_prod}, \eqref{Theta_identity} and then use
    \be
    \Theta (\tau) \delta(\tau) =\frac{1}{2}\delta(\tau)\,.\label{delta_theta_prescr}
\ee
\item  \label{prescript_2} As a second alternative, when evaluating the equations of motion, we retain the powers of $\Theta$
\begin{align}
   & {\Theta^n(\tau)\, , \quad\Theta^n(-\tau)}=[1-\Theta(\tau)]^n\,,\nonumber\\
   &\Theta^j(\tau) \Theta^k(-\tau)=\Theta^j(\tau)[1- \Theta(\tau)]^k\,,
\end{align} and invoke the following rule:
\begin{align}
    \Theta^n(\tau) \delta(\tau)=\frac{1}{n+1}\delta(\tau)\,.
\end{align}
\end{enumerate}

 In both prescriptions, we deduce the following distributional form of the generalized Einstein tensor $\mathcal{E}_{ab}$:
\begin{align}
    \mathcal{E}_{\mu \nu }=& \mathcal{E}^+_{\mu \nu }\Theta(\tau) + \mathcal{E}^-_{\mu \nu }\Theta(-\tau)+ (-k_\rho u^\rho)^{-1}t_{\mu\nu} \delta(\tau)\,,\nonumber\\
    \mathcal{E}_{ij }=& \mathcal{E}^+_{ij }\Theta(\tau) + \mathcal{E}^-_{ij }\Theta(-\tau)+ (-k_\rho u^\rho)^{-1}t_{ij} \delta(\tau)\,,
\end{align}
where the singular parts appearing above can be identified as the singular parts of the energy-momentum tensor of a null shell, whose standard definition is given by \cite{poisson2004relativist}
\begin{align}
    t_{ab}=\mu k_a k_b + j_i(k_a e_b^{i}+k_b e_a^{i} )+p\sigma_{ij}e_{a}^ie_{b}^j\,.\label{shell_em_tensor}
\end{align}
Here, $\mu$, $p$, and $j^i$, respectively, denote the mass density, pressure, and current of the shell.

In prescription \ref{prescrip_1}, we read off the mass density, pressure, and current associated with the shell as
\begin{align}
    \mu=&-\frac{(D-2)}{4\varphi^{D-2}}(G_3^+ + G_3^-) [N^\mu \partial_\mu \varphi]_{\pm}\,,\label{shell_mass}\\
     p=& -\frac{1}{\varphi^{D-3}}\Big \{\frac{1}{4}(G_3^++ G_3^-)[C_{\lambda \lambda}]_{\pm}\nonumber\\
   &+\frac{1}{2}(G_{3,X}^+Q_{+} +G_{3,X}^-  Q_{-}) [N^\mu \partial_\mu \varphi]_{\pm}\Big\}\,, \label{shell_pressure_1}\\
   j^i=&0\,, \label{shell_current}
\end{align}
where $Q_{\pm}$ are defined by
\begin{align}
   {Q_{\pm}\equiv} (k^\mu k^\nu \nabla_\mu \nabla_\nu \varphi)_{\pm}\,.
\end{align}

In prescription \ref{prescript_2}, we find that the expressions for $\mu$ and $j^i$ remain the same as those in prescription \ref{prescrip_1}. The only change is in $p$, which reads
\begin{align}
       p=& -\frac{1}{\varphi^{D-3}}\Big \{\frac{1}{4}(G_3^++ G_3^-)[C_{\lambda \lambda}]_{\pm}\nonumber\\
   &+\frac{1}{6}\Big(2G_{3,X}^+Q_{+} +2G_{3,X}^-  Q_{-}+ G_{3,X}^+Q_{-} \nonumber\\&+G_{3,X}^-  Q_{+}\Big) [N^\mu \partial_\mu \varphi]_{\pm}\Big\}\,. \label{shell_pressure_2}
\end{align}
We point out that in both prescriptions, the term $\nabla_{[\rho}\partial^\sigma\varphi \nabla_{\sigma]}\partial^\rho \varphi G_{3,X}$ in $\mathcal{E}_{ij }$ potentially produces $\delta^2$. However, the term arises with a vanishing coefficient, thanks to $k^ak_a=k^\mu k_\mu=0$, thus effectively eliminating it. We also note that spherical symmetry has forced the shell current to vanish identically, regardless of the prescription used.

Since we are working with theories invariant under diffeomorphisms, it is necessary that the total energy momentum tensor is conserved, i.e., $\nabla^a \mathcal{E}_{ab}=0$. Hence, we have\footnote{Note that $t_{ab}\nabla^a \delta(\tau)= -(-k_\rho u^\rho)^{-1} t_{ab}k^a\delta'(\tau)=0$, as $k^at_{ab}=0$.}
\begin{align}
    \nabla^a \mathcal{E}_{ab}^+ \Theta(\tau) + \nabla^a \mathcal{E}_{ab}^- \Theta(-\tau) + (-k_d u^d)^{-1} A_b \delta(\tau)=0\,,\label{conservation_1}
\end{align}
where 
\begin{align}
    A_b= -[\mathcal{E}_{ab}k^a]_{\pm}+ \nabla^a t_{ab} + t_{ab}\frac{\nabla^a(k_cu^c)}{(-k_d u^d)}\,.
   \end{align}
Since each of the three pieces in equation \eqref{conservation_1} must vanish independently, we have $A_b=0$. Upon contracting this with $N^b$, and using the form \eqref{shell_em_tensor} together with $j^i=0$, we obtain the following conservation equation for the shell
\begin{align}
     &-k^a\nabla _a\mu -\mu \nabla_ak^a+ \mu k^a \nabla_ak^b N_b\nonumber\\
     &+ p\sigma_{ij}e^i_aN^b\nabla^ae^j_b  - \mu \frac{k^a\nabla_a(k_cu^c)}{(-k_d u^d)}-[\mathcal{E}_{ab}k^a N^b]_{\pm}=0\,. \label{conservation_2}
\end{align}

We see that $\mu$, $p$ and $j^i$ are scalars on the two-dimensional manifold associated with $h_{\mu \nu}$, and so are the quantities in the conservation equation \eqref{conservation_2}. Therefore, although our derivation relies on the continuous coordinates $x^\mu$, these scalars can also be computed in the $x^\mu _{\pm}$ coordinates.

\subsection{Static spherically symmetric case}
In the static spherically symmetric case, upon gauge fixing the dilaton field $\varphi$ to the radial coordinate $r$, one can choose Eddington--Finkelstein-like coordinates \eqref{null_coord_metric} for $h_{\mu \nu}$ on each side of the shell. Furthermore, since vacuum black holes in Birkhoff-QT gravity (as well as the charged Schwarzschild--Tangherlini black hole in GR) are determined by a single metric function, we can set $\chi=0$ and thus have
  \begin{align}
ds^2_{\pm}
= - F_{\pm}(r) dv_{\pm}^2 + 2 dv_{\pm} dr + r^2 d\Omega^2_{D-2}\,,\label{vacuum_plus_minus_metric}
\end{align}
Moreover, note that in this gauge $X = F$ and hence $\psi$ takes the form given by equation \eqref{psi_F_rel}.

The normal and auxiliary null vectors can be chosen as
\begin{align}
    &(k_a dx^{a})_{\pm}=-dv_{\pm}\, \Leftrightarrow (k^a \partial_{a})_{\pm}=-\partial_r, \\
    & (N_{ a}dx^{a})_{\pm}=  - \frac{1}{2}F_{\pm}dv_{\pm}+dr\,.
\end{align}
Then it can be shown that
\begin{align}
    &C_{\lambda \lambda}^+=0=C_{\lambda \lambda}^-\,, \quad Q_+=0=Q_-\,,\\
   & N_{\pm}^\mu \partial_{\mu} \varphi= N_\pm^r=\frac{1}{2} F_{\pm}\,,\\
   &(k^a\nabla_a k_b)_\pm=0\,, \quad (\nabla_ak^a)_\pm =-\frac{D-2}{r}\,.
\end{align}
Therefore, in both prescriptions, the shell is pressure-less, {c.f. Eqs.~\eqref{shell_pressure_1} and \eqref{shell_pressure_2}.} Furthermore, one can choose the parameter $\tau$ to be $\tau(x^\mu)=v-v_0$ for some constant $v_0$, so that 
\begin{align}
    k^a u_a= -1, \quad u^a= \Big(1, \frac{dr(v)}{dv}, 0, \dots,0 \Big)\,.
\end{align}
Thus, we see that the shell conservation equation \eqref{conservation_2} reduces to
\begin{align}
   \partial_r(\mu r^{D-2}) -r^{D-2}[\mathcal{E}_{ab}k^a N^b]_{\pm}=0\,. \label{conservation_3}
\end{align}
For the vacuum case, i.e., $\mathcal{E}_{\pm ab}=0$, equation \eqref{conservation_3} implies that $\mu r^{D-2}$ is independent of $r$, and hence constant. Defining the normalized mass of the shell as
\be 
\mathscr{M}_{\text{shell}}\equiv  \frac{\mu r^{D-2} }{(D-2)}\,,
\ee
we find that the jump in the Misner--Sharp mass across the  shell is fixed by \eqref{shell_mass} as
\begin{align}
     \mathscr{M}_{\text{shell}}=&\frac{1}{2}\{h'(\psi)^+ + h'(\psi)^-\}[M(r)]_{\pm}\,,\label{shell_mass_final}
\end{align}
where we have substituted $G_3$ from \eqref{G_3_defn} and the expression for $F(r)$ in terms of the Misner--Sharp mass from \eqref{MS_mass}. Since the LHS of \eqref{shell_mass_final} is constant for the vacuum case while the quantities on the RHS are, in general, $r$-dependent, the relation \eqref{shell_mass_final} appears to restrict theories and spacetimes that admit spherically symmetric null thin shells.

\section{Static spherically symmetric spacetimes in Lovelock theories}
\label{Love_Lock}

Let us first 
consider examples of static spherically symmetric vacuum solutions in (specific) Lovelock theories of gravity. Since Lovelock theories are a subset of QT gravities, the derivation of the junction conditions above applies.

\subsection{Gauss--Bonnet theory}
\label{Gauss_Bonnet_subsection}
The Gauss--Bonnet theory is defined by 
\be 
h(\psi)= \psi + \alpha \psi^2\,.
\ee
The corresponding solution, obtained using equations \eqref{psi_F_rel} and \eqref{h_psi_soln}, then reads \cite{Boulware:1985wk} (see also \cite{Hervik:2019gly})\footnote{Solving the quadratic equation $\psi + \alpha \psi^2=\frac{2m}{r^{D-1}}$ gives two branches of solutions, namely $F= 1+ \frac{r^2}{2\alpha}\Big[1\pm \sqrt{1+ \frac{8\alpha m}{r^{D-1}}} \Big]$. We work with the branch having the negative sign, as it corresponds to the black hole solution. }
\begin{align}
    F= 1+ \frac{r^2}{2\alpha}\Big[1- \sqrt{1+ \frac{8\alpha m}{r^{D-1}}} \Big]\,,
\end{align}
where $\alpha$ is a positive coupling constant defining the theory, and $m$ is the positive mass parameter, identified with the one in \eqref{h_psi_soln}.

The Misner--Sharp mass and the function $h'(\psi)$ are then given by 
\begin{align}
& M(r) =-\frac{r^{D-1}}{4\alpha}\Big[1- \sqrt{1+ \frac{8\alpha m}{r^{D-1}}} \Big]\,,\label{ms_mass_GB}\\
 &   h'(\psi)=\sqrt{1+\frac{8 \alpha m}{r^{D-1}}}\,. \label{h'_GB}
\end{align}
Therefore, the junction condition \eqref{shell_mass_final} becomes
\begin{align}
     \mathscr{M}_{\text{shell}}=&\frac{1}{2}\Big\{\sqrt{1+\frac{8 \alpha m_+}{r^{D-1}}} + \sqrt{1+\frac{8 \alpha m_-}{r^{D-1}}} \Big\}[M(r)]_{\pm}\,,\label{shell_mass_GB}
\end{align}
which, upon using the expression for the Misner–Sharp mass \eqref{ms_mass_GB}, reduces to
\begin{align}
     \mathscr{M}_{\text{shell}}=m_+-m_-\,.
\end{align}
Hence, the Gauss–Bonnet theory admits null thin shells, with the shell mass coinciding with the GR result.

\subsection{Pure Lovelock theory}
Consider next the pure Lovelock theory of order $K$, defined by \cite{Dadhich:2012cv}
\begin{align}
  &  h(\psi)= \alpha \psi^K\,, \quad K\geq 2\,.
\end{align}
Equations \eqref{psi_F_rel}, \eqref{h_psi_soln} fix $\psi$ and $F$ as \cite{Gannouji:2019gnb}
\begin{align}
&\psi=\Big(\frac{2m}{\alpha r^{D-1}}\Big)^{1/K}\,, \label{psi)cubic_Love}\\
& F= 1- r^2\Big(\frac{2m}{\alpha r^{D-1}}\Big)^{1/K}\,.\label{F_cubic_Love}
\end{align}
The quantities required to evaluate the relation \eqref{shell_mass_final}, namely $h'(\psi)$ and $M(r)$, then read
\begin{align}
  & h'(\psi)= K\alpha \psi^{K-1}= K\alpha^{1/K} \Big(\frac{2m}{r^{D-1}}\Big)^{\frac{K-1}{K}}\,,\label{h'psi_cubic_Love}\\
   & M(r)=\frac{r^{D-1}}{2}\Big(\frac{2m}{\alpha r^{D-1}}\Big)^{1/K}\,.
\end{align}
Therefore, the shell mass is given by
\begin{align}
     \mathscr{M}_{\text{shell}}=\frac{K}{2}(m_+^{\frac{K-1}{K}}+ m_{-}^{\frac{K-1}{K}}) (m_{+}^{1/K}-m_{-}^{1/K})\,,\label{shell_mass_Lovelock}
\end{align}
which is a constant, and hence the shell exists. Note, however, that the shell mass differs from that in GR except for $K=2$, i.e., the Gauss--Bonnet theory. Moreover, if we add an Einstein--Hilbert term to the action and treat the Lovelock term as a small perturbation around GR, then for large $r$ the shell mass must converge to the GR result, while for small $r$ it would be modified by the Lovelock contribution \eqref{shell_mass_Lovelock}. Hence, we can qualitatively argue that Lovelock theories of order $K\geq3$, treated as perturbations to the Einstein--Hilbert term, lead to an $r$-dependent shell mass and therefore suggest that the shell is forbidden, while for $K=2$, as we have already seen in Section \ref{Gauss_Bonnet_subsection}, the shell is consistent.

\subsection{ Fine tuned Lovelock theory}

Consider finally the $K$th-order Lovelock theory, which in the symmetry-reduced form is defined by
\begin{align}
   h(\psi)=\sum_{n=0}^{K}\alpha_{n}\psi^n.
\end{align}
$\alpha_0$ is proportional to the cosmological constant, which we set to zero. On choosing $\alpha_K\equiv \alpha$ arbitrarily, $\alpha_1=1$, and fine-tuning the remaining couplings as follows:
\begin{align}
   & \alpha_n=\alpha A^{K-n} \binom{K}{n}\,, \quad  2\leq n <K\,,\\
   &A\equiv (K\alpha)^{-\frac{1}{K-1}}\,,
\end{align}
the following function forms a solution to the corresponding equations of motion, e.g. \cite{Dolan:2014vba}:
\begin{align}
    F= 1+ r^2A\big( 1- Y^{1/K}\big)\,,\quad Y\equiv1+ \frac{2m}{\alpha A^{K}r^{D-1}}\,.
\end{align}
The function $h'(\psi)$ is given by 
\begin{align}
    h'(\psi)=&\frac{dh}{dr}\Big(\frac{d \psi}{dr}\Big)^{-1}=K \alpha A^{K-1}Y^{\frac{K-1}{K}}\,.\label{h'_fine_Love}
\end{align}

\begin{figure}[H]
\includegraphics[width=0.5\textwidth]{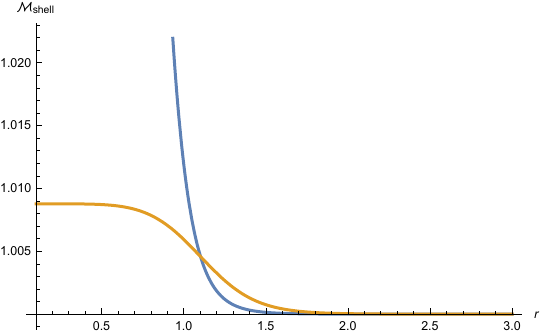}
\caption{{\bf $\mathscr{M}_{\text{shell}}$ for Lovelock and Hayward black holes.} The shell mass $\mathscr{M}_{\text{shell}}$ is displayed as a function of $r$ for fine tuned Lovelock gravity with $K=3$ (orange) and the Hayward black hole (blue). In both cases we choose $D=7$, $\alpha=0.1$, $m_+=2$, and $m_-=1$. Obviously, in both cases the shell mass is not a constant, showing that 
our junction condition forbids 
spherical null shells in these theories. 
} 
\label{fig:plot}
\end{figure}

The Misner--Sharp mass reads
\begin{align}
    M(r)=\frac{A}{2}r^{D-1}\big( Y^{1/K}-1 \big)\,. \label{msmass_fine_Love}
\end{align}
Therefore, the shell mass becomes
\begin{align}
    \mathscr{M}_{\text{shell}}=\frac{K\alpha A^{K}r^{D-1}}{4}\Big (Y_+^{\frac{K-1}{K}}+Y_-^{\frac{K-1}{K}} \Big)\Big ( Y_+^{1/k}- Y_-^{1/k} \Big).
\end{align}
Note that for $K=2$, i.e., the Gauss–Bonnet theory, the expression reduces to the result derived in Sec.~\ref{Gauss_Bonnet_subsection}, giving a constant $ \mathscr{M}_{\text{shell}}$. However, for $K>2$, $ \mathscr{M}_{\text{shell}}$ is not guaranteed to be so.
If we take, for example, $K=3$, $D=7$, $\alpha=0.1$, $m_+=2$, and $m_-=1$, we find that $ \mathscr{M}_{\text{shell}}$ varies with $r$ as shown in Fig. \ref{fig:plot}, and hence is not constant. Therefore, null thin shells need not exist in a generic Lovelock theory.

\section{Applications to vacuum regular black holes in QT gravity}\label{Sec:4}

In this section, we will apply our formula \eqref{shell_mass_final} to 
regular black holes in  QT gravity.

\subsection{Hayward black hole}
\label{Hayward_subsection}
For the Hayward black hole, the functions $h(\psi)$ and $F$ are given by \cite{Bueno:2024dgm}
\begin{align}
    & h(\psi)=\frac{\psi}{1-\alpha \psi}\,,\label{h_psi_Hayward}\\
    & F=1-\frac{2mr^2}{r^{D-1}+2m\alpha}\,.\label{F_Hayward}
\end{align}
 From equation \eqref{h_psi_Hayward}, we find
\begin{align}
    h'(\psi)=\frac{1}{(1-\alpha\psi)^2}\,,
\end{align}
which can be rewritten using equations \eqref{psi_F_rel} and \eqref{F_Hayward} as
\begin{align}
   h'(\psi)= \frac{(r^{D-1}+2m\alpha)^2}{r^{2D-2}}\,.\label{h'_Hayward}
\end{align}
Therefore, from \eqref{shell_mass_final}, we find that the mass of a spherical null shell in the Hayward black hole is given by
\begin{align}
     \mathscr{M}_{\text{shell}}=&\frac{1}{2r^{D-1}}\Bigl((r^{D-1}+2m_+\alpha)^2 + (r^{D-1}+2m_-\alpha)^2\Bigr)\nonumber\\
     &\times \Big[\frac{m}{r^{D-1}+2m\alpha}\Big]_{\pm}\,.\label{shell_mass_Hayward}
\end{align}

The shell mass  on the left hand side is $r$-independent, and hence a constant which can be fixed by considering the $r\rightarrow \infty$ portion of the trajectory. Therefore we find that 
\begin{align}
    \mathscr{M}_{\text{shell}}=m_+-m_-\,,\label{GR_shell_mass}
\end{align}
which is exactly what one derives in GR for Schwarzschild--Tangherlini black holes \cite{Tangherlini:1963bw}. Hence, we obtain the following constraint to be satisfied for all $r$ along the null trajectory
\begin{align}
   \Big( \frac{r^{D-1}+2m_+\alpha}{r^{D-1}+2m_-\alpha} + \frac{r^{D-1}+2m_-\alpha}{r^{D-1}+2m_+\alpha} -2\Big) (m_+-m_-)=0\,,
\end{align}
which can be simplified to
\begin{align}
    (m_+-m_-)^3=0.
\end{align}
Therefore, the only solution to the constraint is \mbox{$m_+=m_-$}, i.e., $\mathscr{M}_{\text{shell}}=0$, and hence spherically symmetric null thin shells are forbidden in the Hayward black hole spacetime.

\subsection{A more general model}
\label{Bardeen_subsection}

Consider the following model, defined by \cite{Bueno:2024dgm,Hennigar:2025yqm}
\begin{align}
   & h(\psi)=\frac{\psi}{(1-\alpha^N\psi^N)^{1/N}}\,,\label{h_Bardeen}
  \end{align}
where $N$ is a natural number. For $N=1$, it reduces to the Hayward black hole, and for $N=2$, it corresponds to a Bardeen-like black hole \cite{1968qtr..conf...87B}. Using equations \eqref{psi_F_rel}, \eqref{h_psi_soln}, one can deduce that
\begin{align}
     &F(r)= 1-\frac{2mr^2}{[r^{(D-1)N}+(2\alpha m)^N]^{1/N}}\,. \label{F_Bardeen}
\end{align}

The expression for $h'(\psi)$ in this case is given by
\begin{align}
    h'(\psi)=\frac{1}{(1-\alpha^N \psi^N)^{\frac{N+1}{N}}}\,. \label{h_prime_Bardeen}
\end{align}
Using equations \eqref{psi_F_rel} and \eqref{F_Bardeen}, $h'(\psi)$ can be rewritten as
\begin{align}
     h'(\psi)=\frac{[r^{(D-1)N}+(2m\alpha)^N]^{\frac{N+1}{N}}}{r^{(N+1)(D-1)}}\,.  \label{h'_Misner_Bardeen}
\end{align}

As in the Hayward case, the constant shell mass is fixed by its value at $r\rightarrow\infty$, i.e., the GR value \eqref{GR_shell_mass}. Hence, substituting \eqref{h'_Misner_Bardeen} and the expression for the Misner--Sharp mass into \eqref{shell_mass_final}, and expanding in $\alpha$, for $N>1$, one finds that the leading terms cancel on both sides, and thus the sub-leading terms, which are of order $\alpha^N$, must vanish on their own, leading to the constraint
\begin{align}
  (N+1)(m_+^N+m_-^N)(m_+-m_-)-2(m_+^{N+1}-m_-^{N+1})=0\,.\label{Bardeen_constr_1}
\end{align}
Using the identity
\begin{align}
    a^{N+1}-b^{N+1}=(a-b) \sum_{k=0}^{N}a^{N-k}b^k\,,
\end{align}
one can rewrite \eqref{Bardeen_constr_1} as
\begin{align}
    (m_+-m_-) \Big[ (N-1)(m_+^N+m_-^N)- 2\sum_{k=1}^{N-1}m_+^{N-k}m_-^k \Big]=0\,.\label{Bardeen_constr_2}
\end{align}
It is easy to see that the expression in the square brackets, as well as its prefactor, vanish for $m_+=m_-$. To show that the expression in square brackets does not admit any other solution, it is sufficient to prove that, for \mbox{$z\equiv \frac{m_+}{m_-}\neq 1$}, the following function is strictly positive:
\begin{align}
H_N(z)= (N-1) (1+ z^N)-2\sum_{k=1}^{N-1}z^{N-k}\,.
\end{align}
We prove this by induction. Note also that $m_+=0$ implies $m_-=0$ and vice versa, and hence we can safely assume $z>0$.

We see that $H_2(z)=(z-1)^2$, which is strictly positive for $z\neq 1$. Assume that $H_N(z)>0$, and consider
\begin{align}
    H_{N+1}(z)= N(1+z^{N+1})-2\sum_{k=1}^{N}z^{N+1-k}\,,
\end{align}
which can be split as
\begin{align}
       H_{N+1}(z)= H_{N}(z)+ \tilde H_N(z)\,,
\end{align}
where $\tilde H_N(z)= 1+ Nz^{N+1}-(N+1)z^N$. It is easy to check that the only extremum of $\tilde H_N(z)$ for $z>0$ is \mbox{$z=1$}. Moreover, $\tilde H_N(1)=0$ and $\tilde H_M''(1)=N(N+1)>0$. Therefore, for $0<z\neq 1$, $\tilde H_N(z)$ is strictly positive, and hence $H_{N+1}(z)$ is strictly positive, concluding our proof.

Thus, we see that, for any $N>1$, $m_+=m_-$ is the only solution of the constraint \eqref{Bardeen_constr_1}, which implies that null shells are ruled out in these cases as well.

\section{Conclusions}\label{Sec:5}

Thin shells are an idealized matter distribution whose energy-momentum tensor is defined in a distributional sense. Unlike in GR, in a generic theory of gravity such a tensor need not admit a consistent distributional metric solution, and hence thin shells may not be consistently realized \cite{Senovilla:2026fby}. This question is particularly relevant to the stability of regular black holes, because if null thin shells can be accommodated, generic arguments supplemented by minimal assumptions lead to the mass inflation instability \cite{Carballo-Rubio:2018pmi} (see, however, \cite{Frolov:2026rcm} for a different interpretation of this result in the QT context).

In this paper, we have derived the junction conditions for spherical null thin shells in QT theories of gravity by employing the distributional method \cite{poisson2004relativist}. In our analysis, we encountered the technical obstacle of $\Theta \delta$ terms, which are distributionally ill-defined. To handle such terms, we employed three different prescriptions, the first being an effective elimination of the product through a continuity relation (see \eqref{cont_prescrpt} and Appendix \ref{Append_1}), in line with standard textbook derivations \cite{poisson2004relativist}, while the other two prescriptions are rules to evaluate the product without eliminating it.

In the prescription \eqref{cont_prescrpt}, which removes the $\Theta \delta$ product, we found that thin shells, including non-null ones, are forbidden in the modified gravity examples we discuss. This is to be contrasted with the
results of e.g. \cite{Davis:2002gn,  Bueno:2024zsx}, where the junction conditions for non-null
shells are derived from the action, and lead to a different conclusion than ours, indicating that the prescription \eqref{cont_prescrpt} is too restrictive.

In the other two prescriptions, which eventually turn out to be equivalent for static spacetimes, we obtained the constraint \eqref{shell_mass_final} as the junction condition fixing the mass of the shell, with its pressure and current found to vanish. The constraint, when applied to GR (see Appendix \ref{mass_inflation_RN_app}), reproduces known results in the literature. In addition, it also allows for the existence of null thin shells in Gauss–Bonnet and pure Lovelock theories, as shown in Sec. \ref{Love_Lock}, indicating that these prescriptions may be reasonable.

We therefore proceeded and applied the junction condition \eqref{shell_mass_final} to the vacuum regular black hole examples considered in Sec. \ref{Sec:4} to examine whether they suffer from mass inflation instability. We found that the junction condition forbids the existence of null thin shells in these regular black hole spacetimes. In fact, as indicated by the last example in Section \ref{Love_Lock}, this is already the case for black holes of more general Lovelock theories.

The results of this work open interesting possibilities for future investigations. 
For example, is it possible to refine the derivation of null junction conditions, using e.g. the regularization methods developed in \cite{Chu:2021uec},  so that these would allow for the existence of null shells in general Lovelock, QT, and non-polynomial QT \cite{Colleaux:2017ibe,Colleaux:2019ckh,Bueno:2025zaj, Borissova:2026wmn} gravities? On the other hand, if the  null thin shells are truly absent in these theories,
it follows that 
the stability analysis of regular black holes in QT  theories is more subtle than previously assumed, e.g. \cite{Frolov:2026rcm}. In particular, the standard double-shell formalism based on the thin-shell approximation cannot be directly applied, and the stability of regular black holes constructed within this framework must therefore be reassessed using a more refined analysis that does not rely on a distributional description of null matter shells.

\subsection*{Acknowledgments}
We would like to thank Ujjwal Agarwal, Raúl Carballo-Rubio and Robie Hennigar for discussions.
F.D.F. acknowledges financial support from the von Humboldt Foundation.
D.K. and A.S. acknowledge support from
 Charles University Research Center Grant No.
UNCE24/SCI/016. A.S. additionally acknowledges support from GAČR 25-15544S grant of the Czech Science Foundation and the research plan RVO 67985840 of the Institute of Mathematics, Czech Academy of Sciences.

\appendix

\section{Comments on the continuity prescription for removing the $\delta\Theta$ product}
\label{Append_1}

In Sec. \ref{subsection_second_jn}, although our discussion focused on null shells, it is easy to see that $\delta \Theta$ terms in the $\mathcal{E}_{\mu \nu}$ equation also arise for non-null shells. These terms originate only from $G_3 h_{\mu [\nu} \nabla_{\beta]}\partial^\beta \varphi$. The double derivative of the dilaton field produces the $\delta$, while the distributional form of $G_{3}$ contributes factors of $\Theta$. Therefore, a way to effectively avoid this product is to demand $G_3^+ = G_3^-$, which, using \eqref{G_3_defn} and the continuity of $\varphi$, implies
\begin{align}
    h'(\psi^+)=h'(\psi^-)\,.\label{h'_cont}
\end{align}

Similarly, if one analyzes the $\mathcal{E}_{ij}$ equations, one would obtain further continuity conditions. However, for the examples discussed in Sections \ref{Love_Lock} and \ref{Sec:4}, the continuity condition \eqref{h'_cont} already proves to be too restrictive. Using their respective expressions for $h'(\psi)$ in Eddington-Finkelstein-like coordinates, and the fact that $\alpha$ is fixed by the theory and hence $\alpha_+=\alpha_-$,
one obtains the following condition in each of the cases
\begin{align}
 m_+ = m_-.\label{M_cont}
\end{align}
Therefore, we also have
\begin{align}
 F_+=F_- \,.\label{F_cont}
\end{align}
It then follows that the derivatives of $F(r)$ must also match on both sides. As $F$ is the only metric function, the shell scalars must be given by the jumps in $F$ and its derivatives, and thus must vanish.\footnote{We see this explicitly for the null shells in each example by substituting the conditions \eqref{h'_cont} and \eqref{F_cont}, together with the corresponding expressions for the Misner--Sharp mass, into \eqref{shell_mass_final}. It is straightforward to verify the analogous result for the non-null case as well. However, the arguments presented here suffice.} Therefore, eliminating the $\Theta \delta$ product forbids the existence of thin shells in all the beyond-GR examples considered in this paper, including the Gauss-Bonnet case discussed in Sec. \ref{Love_Lock}, for which non-null shells are known to exist \cite{Davis:2002gn}, and hence the prescription \eqref{cont_prescrpt} does not appear to be a viable strategy.

\section{Mass inflation in charged Schwarzschild-Tangherlini black hole}\label{mass_inflation_RN_app}

Let us employ the double-shell analysis of Sec. \ref{section_Double_null} and the junction conditions obtained in Sec. \ref{Sec:3} to rederive mass inflation for the charged Schwarzschild-Tangherlini black hole \cite{Tangherlini:1963bw}.

For GR, we have $h(\psi)=\psi$ and $h'(\psi)=1$. Therefore, the expression for $[M]_{\pm}$ in terms of the shell mass, given by \eqref{shell_mass_final}, becomes
\begin{align}
    [M]_{\pm}=\mathscr{M}_{\text{shell}}\,,\label{GR_MS_shell_mass}
\end{align}
which is in agreement with the well-known result for a null spherical shell in static spherically symmetric spacetimes in GR \cite{poisson2004relativist}. For the Schwarzschild--Tangherlini black hole,
the Misner--Sharp mass is given by
\begin{align}
    M(r)=m -\frac{q^2}{2r^{D-3}}\,,
\end{align}
where $q$ and $m$ are, respectively, the charge and mass parameters. Therefore,  assuming that the shell is neutral and hence $q_+=q_-$, the shell mass reads
\begin{align}
    \mathscr{M}_{\text{shell}}=m_+-m_-.
\end{align}

In the double-shell analysis, for the ingoing shell, we identify the region $C$ with $+$ and $B$ with $-$, and hence $\mathscr{M}_{\text{in-shell}}=M_{in}$. Similarly, identifying region $D$ as the corresponding $+$ part and $B$ as the $-$ part, we have $\mathscr{M}_{\text{out-shell}}=M_{out}$. The trajectory of the ingoing shell is taken to be close to that of the ingoing inner horizon, i.e., $(r,v_B)\rightarrow (r_-, \infty)$. Therefore, if we assume it is described by the late-time slice of an appropriate massless matter field--for instance, a minimally coupled massless scalar field--then, from the results of \cite{Cardoso:2003jf,Konoplya:2024hfg}, we can take the temporal behavior of the shell mass to be given by Price's law \cite{Price:1971fb}
\begin{align}
\mathscr{M}_{\text{in-shell}}= \mathscr{M}^0 v_B^{-\gamma}\,,\label{price}
\end{align}
for positive constants $\mathscr{M}^0$ and $\gamma$. On the other hand, the trajectory of the outgoing shell is chosen to lie away from that of the outgoing inner horizon (but inside the outer horizon), so that $\mathscr{M}_{\text{out-shell}}$ is constant.

Thus, the only regions that can acquire a time ($v$) dependence from the null shells are those in the causal future of the ingoing shell, namely the regions $C$ and $A$. Hence, in particular, the Misner--Sharp masses $M_B$ and $M_D$ are constants, and consequently $M_{out}$, defined by \eqref{M_out}, is also constant. The $v$-dependence of $M_{in}$ becomes
\begin{align}
  M_{in}= \mathscr{M}^0 v_B^{-\gamma}\,.\label{time_depend_Min_RN}
\end{align}
Substituting \eqref{time_depend_Min_RN} into \eqref{Mass_rels_2} gives the leading temporal behaviour of $M_A$ as
\begin{align}
      M_{A} =a_1 v_B^{-\gamma}e^{|\kappa_-|v_B}\,,\label{M_A_diverg}
\end{align}
for some constant $a_1$. Thus, we arrive at the well-known mass inflation instability of the charged Schwarzschild--Tangherlini black hole in $D\geq4$.

%



\end{document}